\documentstyle[aps,epsf]{revtex}
\newcommand{\vub}{\mbox{$|V_{ub}|$ }}
\newcommand{\BRbtou}{\mbox{${\cal B}(b\to u l \nu X)$ }}
\newcommand{\btou}{\mbox{$b\to u l \nu X$ }}
\newcommand{\BRbsg}{\mbox{${\cal B}(b\to s\gamma)$ }}
\newcommand{\bsg}{\mbox{$b\to s\gamma$ }}
\begin{document}        
\baselineskip 14pt
\title{Recent Results on Heavy Flavours with ALEPH}
\author{Fabrizio Palla}
\address{INFN sezione di Pisa, Via vecchia livornese 1291, I-56010 S. Piero a Grado (PI) Italy}
%
\maketitle              

\begin{abstract}        

The latest ALEPH measurements on heavy flavours are presented.
In particular the measurements of $|V_{ub}|$, the ${\cal B}(b\to s\gamma)$ and a study on the width difference between mass eigenstates in the $B_s$ system are presented.
\end{abstract}          

\section{Introduction}               

The large sample of $Z$ collected by ALEPH during the LEP1
consists of about 4.5 million hadronic decays. Out of them, about 1 million
of $b\bar b$ events  allow to investigate rare processes which could help
constraining the CKM mixing matrix, before the start of the new experiments at the B factories and hadronic colliders.

In this report I will present the measurements of the \BRbtou \cite{vub} which allows
a measurement of the  $|V_{ub}|$ CKM matrix element, the  ${\cal B}(b\to s\gamma)$ \cite{bsg} and a study on the width difference between mass eigenstates in the $B_s$ system.


\section{Measurement of $|V_{ub}|$}

 The measurement of \vub is performed by measuring the semileptonic branching ratio of B hadrons into charmless final states (\btou). In fact this rate is proportional to the \vub$^2$ and can be computed in the framework of the Heavy Quark Expansion theory \cite{HQET}. 

Previous measurements of \vub were performed in both the exclusive and 
inclusive channels in the $\Upsilon $(4S) \cite{4s}. The exclusive measurements
show large theoretical uncertainties in the transition amplitude computation. 
Inclusive measurements overcame this difficulty looking for an excess of events at the endpoint of the lepton momentum distribution where the $b\to c$ contribution 
vanishes. At LEP both methods are not viable, either due to the small branching ratio  and large background in one case or to the small accuracy in reconstructing the B hadron rest frame in the other case.

This measurement relies upon the fact that nearly 90\% of the \btou decays are  expected to have an invariant mass of the hadronic system $M_X$ below the charm threshold of 1.87 GeV/c$^2$. By contrast, only in 10\% of the cases the lepton energy in the $b$ rest frame, $E^*$, is above the kinematic boundary for $b\to c l \nu X$ transitions \cite{HQET}.

\subsection{Event Selection}

Events are selected by requiring the presence of an identified lepton with momentum larger than 3 GeV/c. In the opposite hemisphere a b lifetime tag is applied to reduce non--b  contamination to less than 2\%. The neutrino energy and direction is estimated from the missing momentum in the lepton hemisphere with a typical accuracy of 280 mrad on its direction and 2 GeV on its energy.

Particles coming from the charmless hadronic system $X$ are reconstructed by means of  two Neural Networks, one to select photons and the other charged particles.  The B hadron rest frame is then reconstructed by adding the momenta of the lepton, the neutrino and the selected particles. The total energy is determined by assigning a mass of 5.38 GeV/c$^2$ to the total system. The momentum and angular resolution, obtained from the simulation, are 4.5 GeV/c and 60 mrad respectively.

The discrimination of the \btou signal events from the background $b\to c$ transitions is made on a statistical basis and is based on the fact that the $c$ quark is heavier with respect to the $u$ quark, leading to different kinematical properties for the two final states. Because of resolution effects the separation that uses a single  kinematical variable such as $M_X$ can be considerably improved by combining more information characterizing the leptonic and the hadronic parts into a multivariate Neural Network.

The choice of the variables is based not only in such a way to have a good discrimination between signal and background, but also on a reduced sensitivity to the specific composition of the $X_u$ system.

Among the Neural Network inputs, the ones with larger separating power are the invariant mass, the sphericity and the lepton momentum. 
The Neural Network output is close to 1 for signal events and close to 0 for background events.

The signal Monte Carlo use a Hybrid model following Ramirez et al. \cite{Ramirez}.
At low hadronic energy (below 1.6 GeV), only resonant states are produced, while  for large energy, non-resonant states are expected to contribute, which occurs in 75\% of the cases.

\begin{figure}[htb]      
\centerline{\epsfxsize 3.8 truein \epsfbox{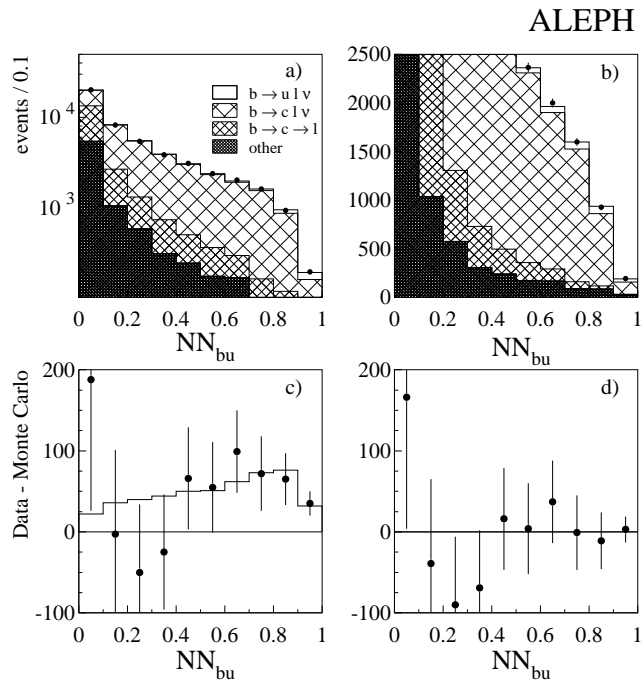}}   
\vskip -.2 cm
\caption[]{\small Neural Net output: a) and b) comparison between data (points) and Monte Carlo (histogram), c) difference between data and Monte Carlo with no $b\to u$ transitions (points) compared to the $b\to u$ contribution (histogram), and d) difference between the data and Monte Carlo with the fitted value of $b\to u$.\label{NNbufitFig}}
\end{figure}

Figure \ref{NNbufitFig} shows the Neural Network output for the Monte Carlo and data after all the selection cuts. The signal is extracted from a binned likelihood fit  to the Neural Network output between 0.6 and 1.0, which gives the smallest total relative error on the measurement. 
The Monte Carlo is normalized to the same number of events in the data to reduce the sensitivity to the assumed efficiencies of the analysis cuts. The first bin is excluded in this normalisation procedure, to minimise the effects of uncertainties of the background events in the fit. In this way, the fit is  effectively a fit to the shape of the signal and background distributions.

\subsection{Results and systematic checks}

An excess of 303 $\pm $ 88 events is clearly seen in the region where the signal is expected, and is in good agreement with the predicted Neural Network shape.

From the fitted number of events one can extract the value of the branching ratio:
\begin{equation}
{\cal B} (b\to u  l  \nu ) = (1.73 \pm 0.55_{\rm stat} \pm 0.51_{{\rm syst }b\to c}\pm 0.21_{{\rm syst }b\to u}) \times 10^{-3}.
\end{equation}

The systematics come from the uncertainties in  modelling  the $b\to c$ and the \btou transitions. For the $b \to c$ systematic source, the main uncertainties are the ones which come from the knowledge of the charm topological branching ratios and the statistical uncertainties on $BR(b\to l)$,
$BR(b\to c\to l )$ and $<X_b>$ (the average energy fraction of the B hadron)  which have been  measured by ALEPH\cite{ALEPH_btol}. The systematic coming from the modelling of the inclusive decays of the $b\to u$ transitions dominates the second systematic error.

Extensive checks have been made to support the measurement. The fit is stable against variations on the fit interval of the Neural Network output, as well as  changing some variables in input of the Neural Network.

Since neutral hadrons have not been considered when reconstructing the $b$ hadron, a bad simulation of the $b\to c$ states involving energetic neutral hadrons would alter the background in the region of high values of
the Neural Network output. 
The neutral hadronic energy distribution in a 30$^0$ cone around the lepton is different for final states with and without $K^0_L$, allowing a measurement of the inclusive production rate of $K^0_L$ in $D$ meson decays. Good agreement is observed between data and simulation. The rate found is in very good agreement with the one measured by MARKIII.

\begin{figure}[htb]      
\centerline{\epsfxsize 2.5 truein \epsfbox{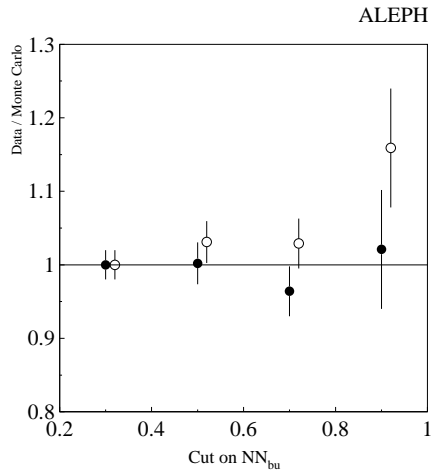}}   
\caption[]{
\small Ratio of the vertexing efficiencies between data and Monte Carlo as a function of the Neural Net output cut. Black circles are with $b\to u$ transitions, while open circles without. \label{vertex}}
\end{figure}

Finally, since no vertexing information has been used in the selection, a common vertex between the lepton and the charged hadronic system is searched for, with a cut on the $\chi^2$ probability larger than 0.2.
The efficiency for this cut is larger for $b\to u$ transitions with respect to the $b\to c$, due to the presence of the charm vertex in the latter. This difference is further enhanced in the Neural Network region close to 1, due to the smaller charged multiplicity and poor vertex of the $b\to c$ transitions. The good agreement of vertexing efficiencies in data and Monte Carlo is shown in Fig.\ref{vertex}, where the effect of the signal inclusion is evident.

From this measurement and the average b lifetime one can extract the value of the \vub matrix element, by using the relation obtained in the framework of the Heavy Quark Expansion Theory \cite{HQET}:

\begin{equation}
|V_{ub}|^2= 20.98 \slantfrac{{\cal B} (b\to X_u  l  \nu_l )}{0.002}\slantfrac{{\rm 1.6 ps}}{\tau_B} (1\pm 0.05_{pert}\pm 0.06_{m_b})\times 10^{-6}
\end{equation}
where $\tau_b=(1.554\pm 0.013)$ ps is the average $b$ hadron lifetime:

\begin{equation}
|V_{ub}|^2=(18.68\pm 5.94_{\rm stat}\pm 5.94_{\rm syst} \pm 1.45_{HQE})\times 10^{-6}.
\end{equation}
corresponding to \vub = $(4.16\pm 1.02)\times 10^{-3}$.

\section{Measurement of the \BRbsg}

The \bsg decay is a flavour changing neutral current that in the Standard Model proceeds via an electromagnetic penguin diagram in which the photon is radiated from either the W or one of the quark lines and its branching ratio is predicted to be $(3.76\pm 0.30)\times 10^{-4}$ at the Z peak.

Virtual particles in the loop may be replaced by non-Standard Model particles, such as charged Higgs bosons or supersymmetric particles which could either enhance or suppress the decay rate, making it sensitive to physics beyond the Standard Model \cite{New_physics}.

The \bsg decay is expected to be dominated by two body decays, leading to the presence of a very energetic photon in the final state associated with a system of high momentum and rapidity hadrons, originating from a displaced secondary vertex.

The composition of the signal \bsg Monte Carlo is based primarily on predictions from the Heavy Quark Effective Theory as well as on the CLEO measurements of the inclusive \bsg branching ratio and the exclusive $B\to K^*(892)\gamma$ \cite{CLEO_bsg}.

\subsection{Event Selection}

The event reconstruction starts from the requirement of a photon with an energy larger than 10 GeV which does not gives any $\pi^0$ when paired with other photons in the event. In that hemisphere, $\pi^0$ and $K_s$ mesons are searched for. The remaining tracks were assigned a probability to come from the \bsg according to their momentum, rapidity with respect to the B hadron direction and, in the case of charged tracks, their three dimensional impact parameter significance. The probability functions were derived from the simulation.

The reconstruction of the \bsg candidate is then made by the photon and the hadronic system in the same hemisphere resulting by adding the  $\pi^0$ and $K_s$, charged tracks and neutral calorimetric objects in decreasing \bsg probability. The candidate is accepted if the jet mass lies within 700 MeV/c$^2$ of the mean B meson mass, the hadronic system has a mass  smaller than 4 GeV/c$^2$ and a multiplicity smaller than eight objects. 

The hemisphere opposite to the candidate is required to be b-like using a lifetime b tag. 

After the preselection there remain 1560 hadronic events, which are then splitted into eight different categories depending on :

\begin{itemize}

\item the value of the length of the major axis of the shower ellipse, $\sigma_l$, of the photon candidate in the electromagnetic calorimeter
\item the energy of the hard photon, $E^*_\gamma$
\item the b tag probability of the opposite hemisphere.

\end{itemize}

The signal is extracted from a binned log-likelihood fit of the  $E^*_\gamma$ distributions of the eight subsamples, using the corresponding distributions for the signal and background simulations, taking into account the finite Monte Carlo statistics in each bin. The five parameters in the fit are $N_{b\to s\gamma}, N_{FSR}, N_{(b\to c)\pi^0}, N_{({\rm non-b})\pi^0}$ and $N_{other}$, which are, respectively, the total number of signal, Final State Radiation, $ (b\to c)\pi^0,({\rm non-b})\pi^0$ and 'other' background events.

\begin{figure}[ht]      
\centerline{\epsfxsize 3. truein \epsfbox{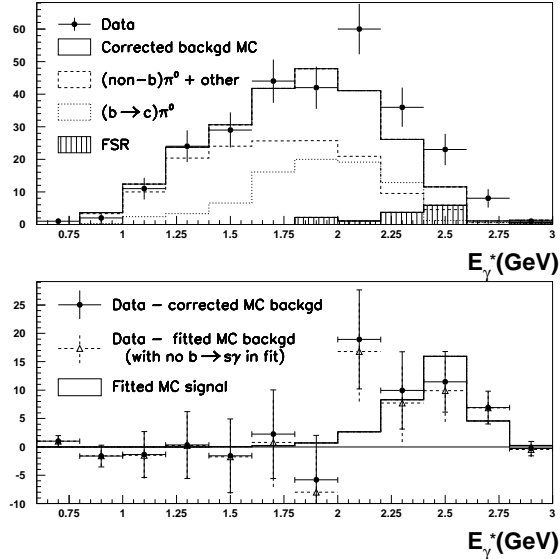}}   
\caption[]{\small  The energy of the photon in the rest frame of the reconstructed jet. The top figure shows the data and the Monte Carlo background events in the region of high purity signal. The bottom figure shows the excess in data after 
the subtraction of the Monte Carlo background and the signal distribution from the fit.
\label{bsgsignal} }
\end{figure}

Figure \ref{bsgsignal} shows the $E^*_\gamma$ distribution for the purest signal sub-sample in data and
Monte Carlo. The clear excess in data  is nicely explained by the signal contribution.

\subsection{Results and systematic checks}

From the number of fitted events and correcting for the  efficiency in the selection cuts the inclusive \bsg
branching ratio is determined to be

\begin{equation}
{\cal B}(b\to s\gamma) = (3.11\pm 0.80_{\rm stat}\pm 0.72_{\rm syst})\times 10^{-4}
\end{equation}
where the statistical error takes into account also the finite Monte Carlo statistics.
The largest systematic error comes from the shape of the  $E^*_\gamma$ background distributions as well as
on the relative proportion of each background source in each sub-sample. This is assessed by observing the
change in the measured branching ratio as the boundaries between the eight sub-classes are varied.
The uncertainty coming from the  energy calibration has been estimated by varying both the ECAL and HCAL 
calibrations. 
Finally the uncertainty of the baryonic \bsg decays has been assessed by repeating the fit setting that fraction to zero.

Checks have been performed which are consistent with the expected \bsg hypothesis:
\begin{itemize}
\item there is evidence of lifetime in the same hemisphere as the photon
\item there is an excess of high momentum kaons 
\item the shape of the excess $\sigma_l$ distribution is characteristic of single photons.
\end{itemize}

\section{A study on the width difference between mass eigenstates in the $B_s$ system}

One of the most challenging measurements in the $B_s-\bar B_s$ system is the
measurement of the mass difference between the two mass eigenstates, $\Delta M_s$, which allows to constrain the CKM unitarity triangle. Since  $\Delta M_s$ is large\cite{PDG} and therefore difficult to measure, a complementary
insight might come from the width difference $\Delta \Gamma_s$, which is related to $\Delta M_s$ via the relation \cite{Voloshin}:
\begin{equation}
\frac{\Delta\Gamma_s}{\Delta M_s} = -\frac{3}{2}\pi \frac{m^2_b}{m^2_t}
  \frac{\eta_{QCD}^{\Delta\Gamma_s}}{\eta_{QCD}^{\Delta M_s}}
\end{equation}
where the ratio of the QCD correction factors ($\eta$) is expected to be of the order of unity, and does not depend on the CKM matrix elements \cite{Dunietz}.
The width difference is expected to be of the order of 10\% of the total width.

One of the simplest way to investigate the width difference is to measure
directly one of the two components of the $B_s$ lifetime. The decay $B_s\to D_s^{(*)+} \bar D_s^{(*)-}X$ is dominantly CP even \cite{Aleksan} and, if one neglects CP violating effects, is the short lifetime eigenstate.

The investigation is performed in the case when each of the two $D_s$ decay into a $\phi$. The $\phi$ are identified in the $K^+K^-$ decay mode.

\subsection{Event selection}

ALEPH has undergone a refinement of its tracking performances of the data already taken at LEP1. Firstly a better pattern recognition in the Silicon Vertex Detector allows to increase the vertexing efficiency of hadronic B decays of about 30\%. In addition, thanks to the readout of the specific ionization induced by charged particles on the TPC pads, the dE/dx measurement efficiency has now increased from 85\% to nearly 100\%, without degrading very much the purity. This measurement is the first one which benefits from this upgrade.

Each kaon candidate is required to have at least 1.5 GeV/c momentum and is identified by requiring the dE/dx to be consistent with the kaon hypothesis and vetoing pions. The angle between the two kaons in the $\phi$ rest frame has to be larger than -0.8. The $\phi\phi$ system is required to have at least 10 GeV/c in momentum and an invariant mass between 2.0 and 4.5 GeV/c$^2$. Most of the combinatorial and fragmentation background is thus removed.

Besides the signal events there remain other sources of double $\phi$ events:
one or both $\phi$ can originate from fragmentation or combinatorial 
background. Three  physics background are also present in $Z\to b\bar b$ and 
$Z\to c\bar c$ events:
\begin{enumerate}
\item $B \to D D_s(X)$, $D_s \to \phi_1 X $,  $D \to \phi_2 X $   
\item $B_{(s)} \to D_{(s)} X$ , $D_{(s)} \to \phi_1 X $ 
and $\phi_2$ from fragmentation 
\item $D_{(s)} \to \phi_1 X$ and $\phi_2$ from fragmentation  
\end{enumerate}
The first reproduces exactly the signal signature and can not be
removed in the selection. However the $D$ decay is Cabibbo suppressed
with respect to the signal process ($D_s \to \phi X \sim 18\%$, 
$D^+ \to \phi X \sim 2.9\%$, $D^0 \to \phi X \sim 0.9\%$).

To have a good tracking quality, each kaon must have at least one hit
in the VDET.
The two $\phi$ are then constrained to a common vertex with
the $\chi^2$ probability of the vertex greater than $1\%$.
At this point it is possible to reconstruct the $B_s$ decay length
as the distance of the $\phi\phi$ vertex from the primary vertex projected along  the momentum direction.
To reject  non $b$ events a b-tag is also demanded.

After this chain of cuts, the contribution due to 
$\phi$ from fragmentation is negligible.

The global efficiency is about $8\%$ with a b-purity of $84\%$
in the final sample.

\subsection{Results}

The $B_s$ lifetime is determined from the proper decay time distribution
of the $\phi\phi$ events. 
For each   $B_s$ candidate,
the proper time is obtained from the decay length
$l$ of the $\phi\phi$  system and the  $B_s$ boost.
The decay length is measured in three dimensions by projecting the
vector joining the interaction point and the  $\phi\phi$  decay vertex onto the
direction of flight of the  $\phi\phi$  resultant momentum. 
The typical resolution of the  $\phi\phi$  decay vertex 
along the direction of flight is
$200 \mu$m. The boost of the $B_s$ is computed from the nucleated jet method \cite{jpsi}
starting from the two $\phi$ tracks.

The  $B_s$ is extracted from an unbinned likelihood fit to the proper
time distribution of the $B_s$ candidates. The background events amount to
78\% of the total, and their proper time distribution has been parametrised
from the events in the two $\phi$ sidebands.

\begin{figure}[htb]      
\centerline{\epsfxsize 3. truein \epsfbox  {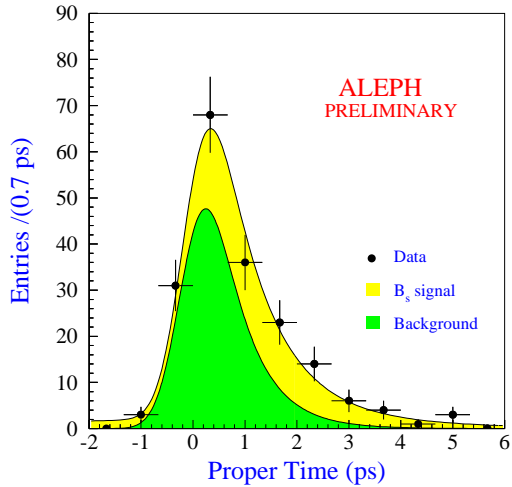}}   
\caption[]{\small Proper time distribution of the two $\phi$ candidates for events selected in the data.
\label{bs_lifetime}
}
\end{figure}

Figure \ref{bs_lifetime} shows the result of the fit:
\begin{equation}
  \tau^{short}_{B_s}= (1.42\pm 0.23 \pm 0.16) {\rm ps.}
\end{equation}

The main systematics comes from  the combinatorial background shape parametrisation.

From this measurement one could extract the difference in the width of the two  $B_s$ mass eigenstates:

\begin{equation}
  \frac{\Delta\Gamma}{\Gamma} = 2 \frac{1-\tau^{short}_{B_s}}{\bar\tau_{B_s}} =
0.24\pm 0.35
\end{equation}
where $\bar\tau_{B_s}$ is the average $B_s$ lifetime (1.61$\pm$ 0.10 ps) and the statistical and systematic errors are combined.

\end{document}